\documentclass[pra,aps,preprint,%showpacs
]{revtex4}
\usepackage{bm}
\usepackage{epsfig}
\usepackage{amssymb}

\input{epsf}

\begin{document}
%\draft

\title{Fractal States in Quantum Information Processing\\}

\author{Gregg Jaeger\\}
\vskip 0.5pt
\affiliation{College of General Studies}
\affiliation{Quantum Imaging Laboratory, \\
Boston University, Boston 02215}

\date{\today}

\begin{abstract}
The fractal character of some quantum properties has been shown for
systems described by continuous variables. Here, a definition of
quantum fractal states is given that suits the discrete systems used
in quantum information processing, including quantum coding and
quantum computing. Several important examples are provided.
\end{abstract}
\maketitle \vfill\eject

\section{Introduction.}

The fractal character of some properties of quantum systems has
often been noted, usually for continuous properties and mainly in
connection with specific spatial, temporal or spectral distributions
-- for example, see \cite{akai,roman,ketzmerick,barboux}.  Here, the
fractal character of discrete quantum systems is considered in the
context of quantum information processing, where quantum parallelism
naturally gives rise to states with very large numbers of
computational basis components. Recently, a specific such class of
self-similar entangled quantum states known as Bell gem states was
defined and shown to include states useful for quantum error
correction and quantum computation \cite{J}. Here, a more general
class of self-similar quantum states, which we will refer to as
``fractal states" that includes Bell gem states, is introduced. Like
geometric fractals, these states are characterized by
self-similarity and potentially fractional values of topological
dimensionality. In addition to showing how quantum concatenation
coding can produce fractal states, other states known to be useful
for quantum information processing are shown to be elements of
sequences of states that are fractal in character when extended to
the limiting scale.

Let us begin by recalling the properties of fractals in general, in
order motivate the definition of quantum fractal states. A fractal
is, broadly speaking, a self-similar entity ({\it e.g.} a set)
having some property that becomes infinite (say, the number of
subentities, {\it e.g.} subsets) and another property that remains
finite ({\it e.g.}, a volume in which it lies) in the limiting scale
\cite{M}. The common feature of fractals is that some property ({\it
e.g.} a symmetry it possesses) is retained with change of scale.
Fractals may be exactly self-similar, approximately self-similar, or
stochastically self-similar. One can also view fractals as limiting
elements of sequences of entities.

The fractals most often considered are geometric sets, exemplified
by the Cantor set  --- see Fig. 1. This set can be viewed as derived
from the interval $[0,1)\in\mathbb{R}$ by successive subdivisions
into pairs of sub-intervals of equal length, one located to the left
and one to the right, with the central interval being excluded. In
the first subdivision of the unit interval, $[0,1)$, the successor
element in the sequence of sets leading to the Cantor set is thus
the set formed by set-theoretic union of the remaining intervals,
$[0,1/3)\cup[2/3, 1)$; the successor to this set is one in which
both these intervals are similarly divided, and so on. The
properties of fractal geometric sets allow one to attribute them a
well-defined topological (``fractal'') dimension $d$, via the
relation $c=s^d$ where $c$ is the number of subdivisions of every
component with each change of scale by a factor $s$, so that $d=\ln
c/\ln s$. For the Cantor set, for example, $c=2$ and $s=3$, so that
$d=\ln 2/\ln 3$, a non-integral real number. Fractals are thus
capable of possessing non-integral topological dimensionality. It is
important to note, however, $d$ may take integral as well as
fractional values \cite{Ft}.

Now let us define quantum fractal states for quantum systems having
discrete states such as might be used in quantum information
applications. This can be done by making correspondences between
quantum state-vector properties and those of geometric sets,
motivating our definition. In particular, we take vector addition in
Hilbert space to play a role analogous to that played by
set-theoretic union in the case of geometric sets. The ratios of
probabilities of outcomes of precise quantum measurements at
successive scales can then be taken to correspond to ratios of
lengths of geometric sets, providing a scaling factor, $s$, for the
fractal. These probabilities are given by squared magnitudes of
(complex) quantum amplitudes of the components of the quantum state
under consideration in the eigenbases determined by precise quantum
measurements. The number of quantum subsystems arising with each
change of scale is finally taken to correspond to the number of
subintervals in the case of geometric sets, providing the parameter
$c$. With these two quantities in hand, one can determine the
fractal dimension of the corresponding quantum fractal state,
specifying its fractal character. One can thus formalize the notion
of a quantum fractal state as in the following definition.

\vskip 20pt

\noindent{\bf Definition:} A {\it quantum fractal state} (QFS) is a
quantum state that is self-similar, having topological
dimensionality $d$ related to the number $c>1$ of subsystems of a
system at each discrete change of scale indexed by $n\in\mathbb{N}$,
and to the scaling factor $s=p^{(n)}/p^{(n+1)}\geq 1$ of
probabilities of measurement outcomes in a given eigenbasis relative
to those at the successor scale:
\begin{equation}
d=\ln c/\ln s
\end{equation}
with $s, c\in \mathbb{N}$, where $d$ is defined to be $d=\log_2 c$
at the bounding scale $s=1$.

\vskip 15pt

In the above, the number of successor subsystems per change of scale
is then $c=s^d$ where, again, the probabilities $p^{(m)}$ are those
obtained from quantum state amplitudes by squaring as per the Born
rule. The following expression provides a specific class of pure
quantum fractal state characterized by the parameters $c$ and $s$,
by relating a term in the sequence of states leading to the fractal
state to its predecessor:
\begin{equation}
|\Phi(n+1,c,s)\rangle= \sum_{i_1,i_2,\ldots ,
i_c=0}^{s-1}\alpha^{(n+1)}_{i_1,i_2,\ldots
i_c}|\Psi_{i_1}\rangle|\Psi_{i_2}\rangle\ldots|\Psi_{i_c}\rangle\ ,
\end{equation}
where $|\Psi_{i_j}\rangle$ are normalized basis vectors for
$\mathcal{H}^{(n)}=\mathcal{H}^{\otimes c^n}$, the $c^n$-dimensional
Hilbert space corresponding to a system at scale $n$, with at least
one of the $|\Psi_{i_j}\rangle$ being the predecessor state
$|\Phi(n,c,s)\rangle$ at the previous scale, under the constraints
$|\alpha^{(n+1)}_{i_1,i_2,\ldots i_c}|^2\in\{0,{1\over s}\}$ and
$\sum_{i_1,i_2,\ldots i_c}|\alpha^{(n+1)}_{i_1i_2\ldots i_c}|^2=1$,
ensuring similarity between scales; a state $|\Phi(0,c,s)\rangle$ at
the initial scale and a basis for Hilbert space $\mathcal{H}$ of
which it is an element must also be identified. Simple, but by no
means necessary, choices in the quantum information processing
setting are the computational basis state
$|\Phi(0,c,s)\rangle=|0\rangle$ and the qu-$N$-it space
$\mathcal{H}=\mathbb{C}^N$, $N\in\mathbb{N}$. To find a
representative fractal state, $|\mathcal{F}(c,s)\rangle$, for a
given set of parameters $(c,s)$, it suffices to take
$|\Psi_{i_1}\rangle=|\Phi(0,c,s)\rangle$,
$|\Psi_{i_j}\rangle=|j\rangle$ for $j>0$,
$\alpha^{(n+1)}_{i_1i_2\ldots i_c}=1/\sqrt{s}$ for $i_1=i_2=\ldots
=i_c$ and $\alpha^{(n+1)}_{i_1i_2\ldots i_c}=0$ otherwise, at all
scales. The states given by Eq. 2 are elements of a sequence leading
to a quantum fractal state in the limit $n\rightarrow\infty$. This
particular state is a quantum product state.

As a specific example in which the above convenient choices are made
that illustrates the fractal character of such states, consider a
quantum state having the characteristic parameters of the Cantor
set, namely $c=2$ and $s=3$, which we will call a ``Cantor state,"
arrived at through the following sequence of states:
\begin{eqnarray}
|\Phi(0,2,3)\rangle&=&|0\rangle\\
|\Phi(1,2,3)\rangle&=&{1\over\sqrt{3}}\bigg(|0\rangle\bigg)\bigg(|0\rangle +|1\rangle +|2\rangle\bigg)\\
&=&{1\over\sqrt{3}}\bigg(|00\rangle +|01\rangle +|02\rangle\bigg)\nonumber\\
|\Phi(2,2,3)\rangle&=&{1\over\sqrt{3}}\bigg[{1\over\sqrt{3}}\bigg(|0\rangle\bigg)\bigg(|0\rangle
+|1\rangle
+|2\rangle\bigg)\bigg]\bigg[|00\rangle +|11\rangle +|22\rangle\bigg]\\
&=&{1\over\sqrt{9}}\bigg(|0000\rangle +|0011\rangle
+|0022\rangle +|0100\rangle +|0111\rangle +|0122\rangle +|0200\rangle +|0211\rangle +|0222\rangle\bigg)\nonumber\\
&\vdots&\nonumber
\end{eqnarray}
where, in particular, the computational basis for $\mathbb{C}^3$ was
chosen. In the limit $n\rightarrow\infty$ this sequence approaches
the representative fractal state $|{\mathcal F}(2,3)\rangle$ having
topological dimension $d=\ln 2/\ln 3$.

As a second example \emph{without} the above convenient choices but
still of the form of Eq. 2, consider Bell gem states of the form
$(1/\sqrt{2})\big(|i\rangle|j\rangle\pm|j\rangle|i\rangle\big)$,
lying in a $d=2^{2^m}$-dimensional Hilbert space, where
$|i\rangle\neq|j\rangle$ are elements of the same form but
dimensionality $d'=2^{2^{(m-1)}}$, $m\in {\mathbb N}, m\geq 2$
\cite{J}; the simplest such states are
\begin{equation}
|\Psi^\pm\rangle={1\over\sqrt{2}}\bigg(|01\rangle\pm|10\rangle\bigg)\
.
\end{equation}
Take as the state at scale $n=0$, $|\Psi^-\rangle$; at the successor
scale $n=1$, one can then have the states
\begin{eqnarray}
{1\over\sqrt{2}}\bigg(|\Psi^+\rangle|\Psi^-\rangle
+|\Psi^-\rangle|\Psi^+\rangle\bigg)&=&{1\over\sqrt{2}}\bigg(|0101\rangle-|1010\rangle\bigg)\\
{1\over\sqrt{2}}\bigg(|\Psi^+\rangle|\Psi^-\rangle
-|\Psi^-\rangle|\Psi^+\rangle\bigg)&=&{1\over\sqrt{2}}\bigg(|1001\rangle-|0110\rangle\bigg)\
,
\end{eqnarray}
having the form of the right-hand side of  Eq. (2) with $c=s=2$,
where in the first case
$\alpha_{01}^{(1)}=\alpha_{10}^{(1)}=1/\sqrt{2}$, all other
coefficients being zero, and in the second case
$\alpha_{01}^{(1)}=1/\sqrt{2}$, $\alpha_{10}^{(1)}=-1/\sqrt{2}$, all
other coefficients being zero. In the limit $n\rightarrow\infty$,
given similarly chosen $\alpha^{(m)}_{ij}$ according with these
symmetries at all scales, one obtains quantum states having
topological dimensionality $d=1$, which are fractals \cite{Ft}.

\section{Relation to computational states}

One can further relate states used in various quantum information
processing applications to quantum fractal states, as we will now
see. Consider quantum code states formed by the continued
concatenation of coding maps. Such states can be viewed as
increasingly well approximating fractal states, in the sense of
being self-similar at an increasing number of scales. The simplest
case of code concatenation involves a single concatenation step: one
uses an $M$-qubit code $C^{\rm out}=({\mathcal E}^{\rm out},
{\mathcal D}^{\rm out})$ referred to as the outer code, and a second
$M'$-qubit code $C^{\rm in}=({\mathcal E}^{\rm in}, {\mathcal
D}^{\rm in})$, referred to as the inner code, where $\mathcal{E}$
and $\mathcal{D}$ indicate encoding and decoding operations, which
are CPTP maps of the statistical operators $\rho$ corresponding to
computational states
--- in our context, $\rho=|\Phi(m,c,s)\rangle\langle\Phi(m,c,s)|$. The
logical qubits of the inner code form $M'$-qubit blocks used by the
outer code, comprising a fully encoded quantum computational
register, in effect using the concatenated encoding map
$\tilde{\mathcal E}={\mathcal E}^{\rm out}\circ({\mathcal E}^{\rm
in})^{\otimes M}$. An error correction scheme can coherently correct
each code block using the inner code and then the entire $MM'$-qubit
register using the outer code \cite{R}. To operate at increasingly
larger scales, a larger concatenation code repeatedly using the same
code for {\it both} inner and outer codes at successive scales can
be used to produce code states. Such states are by definition
quantum fractal states, being of the form given in Eq. (2).

As a specific example, consider the states provided by the bit-flip
code, where the computational basis elements $|0\rangle$,
$|1\rangle$ are mapped onto states of several qubits, forming
logical qubits used for quantum error detection and correction ---
see, for example \cite{BR}. This three-qubit repetition code is
implemented by the following mapping
\begin{eqnarray}
|0\rangle&\mapsto&|0\rangle_L=|000\rangle\\
|1\rangle&\mapsto&|1\rangle_L=|111\rangle\ ,
\end{eqnarray}
where ``$L$'' indicates a logical qubit state. One can then see that
employing a 3-qubit bit-flip code for both inner and outer codes,
allowing for the correction of both physical bit-flip errors and
first-level logical-bit-flip errors, requires a 9-qubit physical
register of three 3-qubit blocks. This quantum bit-flip code, when
continually applied, produces fractal quantum states with parameters
values $c=3$ and $s=1$; the states of the left-hand side of Eqs.
9--10 being those of the scale $0$ and those of the right-hand side
being those of the scale $1$; the single coefficient $\alpha^{(m)}$
at each scale being simply 1. Such an error-correction code can be
used when physical and logical bit-flip errors are possible
simultaneously at different scales. In particular, if a quantum
computational system is found susceptible to logical-bit value
errors at $n$ different scales, the following sequence of
concatenation code states can be used:
\begin{eqnarray}
|{\mathcal
F}(0,3,1)\rangle&=&|i\rangle\\
|{\mathcal
F}(1,3,1)\rangle&=&|i\rangle^{\otimes 3}\\
|{\mathcal
F}(2,3,1)\rangle&=&|i\rangle^{\otimes 9}\\
\vdots\\
|{\mathcal F}(n,3,1)\rangle&=&|i\rangle^{\otimes 3^n}
\end{eqnarray}
with $i=0,1$, which, if continued indefinitely, leads to quantum
fractal states of topological dimensionality $d=\log_2 3$. A series
of $n$ quantum cloning machines each set to output $3$ copies, input
serially, will produce these code states. This is an extremely
simple example. A less simple example of a concatenated quantum
code, based on the encoding map from logic bits to the Bell states
$|\Psi^\pm\rangle$, is that provided as the second example of the
previous section, known to be useful for carrying out quantum error
correction and quantum computation --- see \cite{JS,J} and
references therein.
%Creating such a sequence is seen to
%requires a simple version of the general apparatus provided above
%but where no multiports are needed.

Yet another place where sequences of self-similar states might be
expected to arise in quantum information processing is in spin
clusters --- that is, arrays of qubits. One class of states that has
been proposed for use in cluster quantum computing on a linear qubit
array is given by the expression
\begin{equation}
|\phi_N\rangle={1\over
2^{N/2}}\bigotimes_{a=1}^N\bigg(|0\rangle\sigma_z^{a+1}+|1\rangle\bigg)\
,
\end{equation}
with the convention that $\sigma_z^{N+1}\equiv\mathbb{I}$, which
states are generally considered as given by a representative states
reachable by a local unitary transformation and so are considered
here \cite{BR}. For example, when $N=4$, one has
\begin{equation}
|\phi_4\rangle=_{l.u.t.}{1\over
2}\bigg(|0\rangle|0\rangle|0\rangle|0\rangle
+|0\rangle|0\rangle|1\rangle|1\rangle
+|1\rangle|1\rangle|0\rangle|0\rangle
-|1\rangle|1\rangle|1\rangle|1\rangle\bigg)\ ,
\end{equation}
which can be viewed as of the form $|\Phi(1,2,2)\rangle$, with %${\mathcal
%F}(0,2,2)\rangle=|0\rangle+|1\rangle$ and $u^{(1)}_{i,j}=1$, so that
$|\Phi(0,2,2)\rangle=|\phi_2\rangle=(1/\sqrt{2})(|00\rangle+|11\rangle)$
and
$\alpha^{(1)}_{(00)(00)}=\alpha^{(1)}_{(00)(11)}=\alpha^{(1)}_{(11)(00)}=
(1/\sqrt{2})=-\alpha^{(1)}_{(11)(11)}$. The prescription of Eq. 16
gives rise in the limit of large $N$ to a quantum fractal state much
like the very first example above but with topological
dimensionality $d=1$ since $c=s=2$; the effect of the
$\sigma_z^{a+1}$ is to provide varying phases in the $\alpha^{(N)}$.

Finally, note that the various examples provided above demonstrate
that quantum fractal states need not be specifically only factorable
or only entangled.

\section{Conclusion}

A new class of quantum states was introduced,  the quantum fractal
states, characterized by self-similarity and the capacity for
fractional dimensionality. A specific example class of quantum
fractal states was given and the self-similarity of a number of
quantum states known to be useful for quantum information processing
was pointed out. A stimulus for the broader recognition of
self-similarity in states used for quantum information processing,
where multiple-component systems and quantum error-correction coding
are necessary, has thereby been provided.

\vfill\eject

\centerline{FIGURE CAPTIONS}

\vskip 30pt

{\bf Figure 1.} The Cantor set, lying in the interval
$[0,1]\subset\mathbb{R}$. Dots indicate that the subdivision and
scaling of line segments continues indefinitely.

 \vfill\eject


\begin{thebibliography}{  }
\bibitem{Ft} A common misconception regarding fractals is that $d$
\emph{must be} fractional for them, which is not necessarily the
case, as it is not in this example \cite{M}.
\end{thebibliography}
\end{document}